# Efficient domain wall motion driven by the out-of-plane spin polarization

Jingwei Long[1], Yumeng Yang[1,2], Shuyuan Shi[3], Xuanyao Fong[4], Gengchiau Liang[4], Zhifeng Zhu[1,2†]

1. School of Information Science and Technology, ShanghaiTech University, Shanghai, China 201210
2. Shanghai Engineering Research Center of Energy Efficient and Custom AI IC, Shanghai, China 201210
3. Fert Beijing Institute, MIIT Key Laboratory of Spintronics, School of Integrated Circuit Science and Engineering, Beihang University, Beijing, China
4. Department of Electrical and Computer Engineering, National University of Singapore, Singapore 117576

Fast domain wall motion in systems with perpendicular magnetization is necessary for many novel applications such as the racetrack memory, domain wall logic devices and artificial synapses. The domain wall speed has been greatly improved after the demonstration of current driven domain wall motion using the spin transfer torque, and later achieved another leap due to the combined effect of spin-Hall effect and Dzyaloshinskii-Moriya interaction (DMI). However, only Neel wall can be effectively driven in the latter system. In this paper, we show that both Bloch and Neel domain walls can be effectively moved by exploiting the field-like torque with the out-of-plane spin polarization, which can be generated in materials with reduced symmetry. In addition, we find that the Neel wall and Bloch walls stabilized by the magnetostatic energy are easily distorted by the spin-orbit torque and they then enter the Walker breakdown region. External magnetic field and DMI are introduced to stabilize the domain wall, and a very high domain wall velocity over 500 m/s is obtained at a moderate current density of $2\times10^{11}$ A/m$^2$. In principle, this mechanism can be used to drive any types of domain walls. This work thus provides new degrees of freedom in the control of domain wall motion.

**Key words**

Domain wall motion, out-of-plane spin polarization, field-like torque, Dzyaloshinskii-Moriya interaction

**Introduction**

Since the discovery of current driven domain wall (DW) motion [1], the unidirectional DW motion becomes possible. This produces the idea of racetrack memory [2], which is an innate three-dimensional device and promising for ultradense data storage. The writing performance of racetrack memory depends on the velocity of domain wall motion ($v_{DW}$). Thus, intensive studies have been performed to improve $v_{DW}$. Based on the exchange of angular momentum between the conduction electron and local magnetization, spin transfer torque (STT) enables $v_{DW}$ of 110 m/s [3, 4]. Later on, it is found that the spin-orbit torque (SOT) produced in a heavy metal and a perpendicular ferromagnet heterostructure is able to effectively drive the DW at 400 m/s [5]. To explain this phenomenon, a type of asymmetric exchange interaction, known as the Dzyaloshinskii-Moriya interaction (DMI) [6, 7], has to be taken into consideration. This DMI stabilizes a Neel wall with certain chirality, which can then be effectively moved by the damping-like torque (DLT) produced by the spin-Hall effect (SHE) [8-11]. In addition, the field-like torque (FLT) is demonstrated to have no effect in the DW motion [12]. Therefore, the efficient DW motion driven by SOT is restricted to certain DW profile and driving mechanism, restricting the freedoms in material choice and device design.

In this paper, we show that by exploiting a recent discovered field-like torque with out-of-plane spin polarization ($\tau_{FLT,z} = \mathbf{m} \times \boldsymbol{\sigma}_z$ where $\boldsymbol{\sigma}_z$ denotes the spin polarized in the $\mathbf{z}$ direction) [13-15], both Bloch wall and Neel wall in a perpendicular ferromagnet can be efficiently moved by the current. This can be understood by analyzing the corresponding effective field ($\mathbf{H}_{eff,FLT} = \boldsymbol{\sigma}_z$), which is in the same direction with the neighbor domains. However, when the current density ($J_c$) is large, both Bloch and Neel walls become unstable and enter the Walker breakdown region, resulting in a sharp drop in $v_{DW}$. To stabilize the DW, we firstly apply an external magnetic field ($\mathbf{B}_{ext}$). The requirement of $\mathbf{B}_{ext}$ can be alleviated by utilizing the DMI. As a result, a very high $v_{DW}$ of 506 m/s is obtained at a moderate $J_c$ of $2 \times 10^{11}$ A/m$^2$. These results could stimulate the study of ultrafast DW motion, which is the basis of many novel applications such as the racetrack memory, domain wall logic [16] and artificial synapse [17, 18]. The study presented here also provides insights in designing high performance racetrack memory which is not restricted to the Neel wall driven by SHE.

**Methods**

As shown in Fig. 1, the system under investigation is a nanostrip with perpendicular magnetization in a dimension of 7680 nm $\times$ *width* $\times$ 3 nm, where the width (*w*) is varied to get different DW profiles in equilibrium [19]. For the sample with small width, the magnetostatic energy is mainly contributed by the magnetic surface poles along the **y** direction. Thus, a Neel wall is energetically favorable, whose magnetization is pointed in the **x** direction [see Fig. 1(a)]. However, when the sample width is increased to be larger than the DW width [Fig. 1(b)], most magnetostatic energy cost at the DW surface along the **x** direction, leading to a stable Bloch wall at equilibrium. These predictions are verified by micromagnetic simulations. In samples with different widths, we set the initial DW pointing to 45 degrees in the *xy* plane. It is then relaxed to obtained the equilibrium DW profile. As shown in Fig. 1(d), when the width is smaller than 87.5 nm, the magnetization at the center of DW is pointed in the **x** direction (i.e., Neel wall). In contrast, the Bloch wall is more stable for samples wider than 137.5 nm. Between the two regions, there is a transition region, in which the DW consists of both Bloch and Neel components. This can be easily understood by analyzing the competition of magnetic surface poles along the **x** and **y** directions. Therefore, we choose the width of 50 nm and 400 nm to study the magnetization dynamics of Neel wall and Bloch wall in the following discussions. We have also studied another set of Neel wall (*w* = 30 nm) and Bloch wall (*w* = 300 nm), and qualitatively same results are obtained.

The current driven magnetization dynamics is studied by solving the Landau-Lifshitz-Gilbert (LLG) equation with the addition of $\tau_{FLT,z}$. We have modified the source code of Mumax$^3$ [20] so that the effect of current only produces FLT without invoking the DLT. $\tau_{FLT,z}$ can then be described as

$$\frac{\hbar \theta_{SH} J_c}{2M_s e t_{FL}(1+\alpha^2)}(\mathbf{m}\times\boldsymbol{\sigma}_z + \alpha\mathbf{m}\times(\mathbf{m}\times\boldsymbol{\sigma}_z))$$ after expanding the LLG equation into the Landau-Lifshitz (LL) form [21]. The following parameters are used in the simulations: the anisotropy constant $K_u = 3.3\times10^5$ J/m$^3$ with the easy axis pointing long the **z** axis [12], the damping constant $\alpha = 0.02$, the saturation magnetization $M_s = 6.5\times10^5$ A/m, the exchange stiffness $A_{ex} = 2\times10^{-11}$ J/m and the spin Hall angle $\theta_{SH} = 0.1$. The sample is discretized into cells with the dimension of 3 nm $\times$ 3.125 nm $\times$ 3 nm.

**Results and Discussions**

In the system consists of heavy metal and ferromagnet bilayer, the charge current along the **x** direction can produce vertical spin current $\mathbf{J}_s = \theta_{SH}\boldsymbol{\sigma}\times\mathbf{J}_c$, where $\theta_{SH}$ measures the efficiency of spin current conversion. In this system, the spin polarization is constrained to the **y** direction ($\boldsymbol{\sigma}_y$) due to the rotational symmetry in the film plane [22]. Different mechanisms have been proposed to produce $\boldsymbol{\sigma}_y$, such as the SHE and the Rashba effect. The SHE is predicted to mainly generate DLT and the corresponding effective field ($\mathbf{H}_{eff,DLT}$) has the form of $\mathbf{m}\times\boldsymbol{\sigma}_y$. Since **m** at the center of Neel wall is pointed in the **x** direction, it gives rise to a $\mathbf{H}_{eff,DLT}$ in the **z** direction. Thus, the DW can be effectively moved and $v_{DW}$ is determined by the strength of $\mathbf{H}_{eff,DLT}$ [Fig. 1(c)]. In contrast, the FLT generated by the Rashba effect has the $\mathbf{H}_{eff,FLT}$ in the form of $\boldsymbol{\sigma}_y$, which is pointed in the **y** direction. Thus, it cannot move the DW. In the SHE driven DW motion, it is also worth noting that the magnitude of $\mathbf{H}_{eff,DLT}$ is strongly associated with the $x$ component of magnetization at the DW center ($m_x$). When $m_x$ is reduced during the magnetization dynamics, $\mathbf{H}_{eff,DLT}$ is also reduced, resulting in a smaller $v_{DW}$.

Recently, it has been demonstrated that $\boldsymbol{\sigma}_z$ can be produced in heterostructures with reduced symmetry [13-15, 23-32]. This spin current then exerts torque on the adjacent ferromagnet, and both DLT [$\boldsymbol{\tau}_{DLT,z} = -\mathbf{m}\times(\mathbf{m}\times\boldsymbol{\sigma}_z)$] [24] and FLT ($\boldsymbol{\tau}_{FLT,z} = \mathbf{m}\times\boldsymbol{\sigma}_z$) have been discovered [15, 32]. In these systems, $\mathbf{H}_{eff,DLT} = \mathbf{m}\times\boldsymbol{\sigma}_z$ points in the **y** and **x** directions for the Neel wall and Bloch wall, respectively. Thus, both DWs cannot be driven by the DLT. In contrast, $\mathbf{H}_{eff,FLT}$ is always pointed in the **z** direction. It is independent on the DW profile. As a result, any types of DWs can be driven by $\mathbf{H}_{eff,FLT}$, and we expect a constant $v_{DW}$ even the DW is distorted during the motion. This constancy of $v_{DW}$ is very attractive for the DW based applications.

To verify our proposal, we first study the sample with $w = 50$ nm. After relaxation, a stable Neel wall is obtained. The current is then applied to driven the DW motion. For all the results in this paper, we only consider $\boldsymbol{\tau}_{FLT,z} = -\mathbf{m}\times\boldsymbol{\sigma}_z$. The $\boldsymbol{\tau}_{DLT,z}$, $\boldsymbol{\tau}_{FLT,y}$ and $\boldsymbol{\tau}_{DLT,y}$ are not included. As shown in Fig. 2(a), $v_{DW}$ increases linearly at small current density. However, we find that the DW deviates from the Neel wall and a small $\mathbf{m}_y > 0$ is developed when the current is injected. This can be explained by noting that $\boldsymbol{\tau}_{FLT,z} = -\mathbf{m}\times\boldsymbol{\sigma}_z$, which pulls **m** to the **y** direction. At the same time, **m** experiences an internal torque due to the shape anisotropy, which tries to maintain a Neel wall. Thus, the balance of these two torques results in a stable DW slightly deviated from the perfect Neel wall. As $J_c$ is further increased to $2\times10^{10}$ A/m$^2$, a sudden drop in $v_{DW}$ is observed. To understand this result, the evolution of DW position at $J_c = 3\times10^{10}$ A/m$^2$ is plotted in Fig. 2(b), where an evident back and forth motion is observed. In addition, the DW magnetization rotates along with the DW motion. As shown in Fig. 2(b) using the rectangular symbol, $m_x$ and $m_y$ (not shown) oscillates as a function of time. These two phenomena are typical indications of the Walker breakdown, where the DW does not have a stable profile and thus cannot maintain a uniform motion [33]. In the Walker breakdown region, $v_{DW}$ is calculated as the average velocity in one period. Furthermore, **m** rotates along the counterclockwise direction in our system, which is in consistent with the effect of $\tau_{FLT,z}$.

Next, we study the sample with $w = 400$ nm, which gives rise to a Bloch wall at equilibrium. Similar to the Neel wall, $v_{DW}$ increases linearly at small current density until a Walker breakdown is initiated at $1.8 \times 10^{10}$ A/m$^2$ [see Fig. 3(a)]. This value is slightly smaller than that in the Neel wall. Furthermore, we find that the DW in the Walker breakdown region changes into a vortex, which has not been observed in the Neel wall system. The appearance of vortex in the wide sample has also been observed in the in-plane system [34-36], which can be explained by the different demagnetization energy between the Neel wall and Bloch wall. For the linear region, it is important to note that $v_{DW}$ is almost the same for the Bloch wall and Neel wall [see Fig. 3(b)]. The maximum velocity difference is only 1.7 m/s at $J_c = 0.18 \times 10^{10}$ A/m$^2$, which can be safely ignored. This demonstrates that both DWs can be effectively driven by $\tau_{FLT,z}$. In addition, despite the existence of a small $m_y$ ($m_x$) in the Neel (Bloch) wall, the DWs can still be effectively moved. This demonstrates that the $\tau_{FLT,z}$ is able to drive any types of DWs.

For the DW based applications, a higher $v_{DW}$ is often preferred. Thus, it is important to enlarge the linear $v_{DW}$ region by increasing the Walker breakdown current density. Since the Walker breakdown is induced by the loses of stable DW profile, a straightforward solution is by applying $\mathbf{B}_{ext}$ to maintain the DW profile [35]. Therefore, $H_x$ with different strength is applied to stabilize the Neel wall [see Fig. 4(a)], and all of them show a much larger $v_{DW}$ compared to that in Fig. 2(a). Under $H_x$ = 300 Oe, the maximum $v_{DW}$ is 126 m/s at $J_c = 0.6 \times 10^{11}$ A/m². This velocity is more than three times compared to that without external field (i.e., $v_{DW}$ = 36 m/s at $J_c$ = 0.2 ×10¹¹ A/m²). For $H_x$ = 50 Oe, a drop in $v_{DW}$ is observed when the current density is larger than $0.3 \times 10^{11}$ A/m². Inspecting the DW profile shows that the DW enters the Walker breakdown region. Similar results are observed for $H_x$ = 100 Oe at $J_c = 0.5 \times 10^{11}$ A/m², whereas no Walker breakdown is observed for $H_x$ = 300 Oe. The difference in the breakdown current density demonstrates that the DW profile is stabilized by $H_x$. Next, $H_y$ is applied to stabilize the Bloch wall. As shown in Fig. 4(b), a similar improvement in $v_{DW}$ is observed.

In device applications, $\mathbf{B}_{ext}$ should be avoided since it limits the device scalability and impairs the device performance. To maintain a stable DW without $\mathbf{B}_{ext}$, one can turn to the DMI. Its energy density in micromagnetics is given by $\varepsilon_{DM} = D\left[m_z \frac{\partial m_x}{\partial x} - m_x \frac{\partial m_z}{\partial x}\right] + id.(x \to y)$ [8, 37], where the coefficient $D$ measures the strength of DMI. The energy density indicates that a sufficient large DMI would stabilize a Neel wall whose chirality is determined by the sign of $\mathbf{D}$ [6]. In our system, a positive $\mathbf{D}$ is used and it gives rise to a left-handed Neel wall at equilibrium [37, 38]. The current is then applied to drive the DW motion. As shown in Fig. 5, very fast DW motion has been observed for $D$ = 1.5 mJ/m$^2$. It is worth noting that the maximum $v_{DW}$ exceeds 500 m/s under zero $\mathbf{B}_{ext}$. In addition, even at a large current density of 2× 10$^{11}$ A/m$^2$, the DW maintains a stable profile and no Walker breakdown is observed. At small current density (e.g., $J_c$ < 1×10$^{11}$ A/m$^2$), the DMI strength has almost negligible effect on $v_{DW}$. This is easily understood since the DW motion driven by $\tau_{FLT,z}$ is independent on $\mathbf{m}$ [39]. Thus, under the same current density, although $m_y$ is not the same for samples with different $D$, $v_{DW}$ is not affected. This remains true as long as the DW maintains a stable profile, which is supported by the DMI in our system. Note that this is strikingly different from the motion of Neel wall driven by SHE, where the increase of $m_y$ due the DW distortion reduces $v_{DW}$. However, when the current density is further increased, larger DMI constant leads to faster motion. This is induced by the change of DW width, and a wider DW moves faster [35, 40]. In our system, for example, the DW width for $D$ = 1.5 mJ/m$^2$ is larger than that $D$ = 1 mJ/m$^2$ in at the same $J_c$ of 1.6×10$^{11}$ A/m$^2$. Another interesting results in our system is that the DW tilting is observed after introducing the DMI. Similar DW tilting have been observed in the SHE driven DW motion in systems with perpendicular magnetization [41, 42] or in-plane magnetization [29]. To determine the direction of tilting DW, one can make use of the $m_y$ at the DW center. Since $\tau_{FLT,z}$ is pointed in the $\mathbf{y}$ direction, a small $m_y$ is induced under the current. Since the DMI tends to maintain a Neel wall, the DW magnetization has to be perpendicular to the DW front so that the total energy cost is minimized. In our system, $m_y$ is larger than zero, resulting in a counterclockwise tilting [41, 43].

Finally, we summarize the improvement of $v_{DW}$ in Table 1. In the absence of $\mathbf{B}_{ext}$ and DMI, the DW quickly loses stability, resulting in the maximum $v_{DW}$ of only 36 m/s. By applying a large field of $H_x$ = 300 Oe, the DW can be stabilized and fast DW motion at 126 m/s is obtained. To get rid of the $\mathbf{B}_{ext}$ and maintain a stable DW profile, materials with large DMI can be used and a very high $v_{DW}$ = 506 m/s is achieved.

**Conclusion**

In conclusion, we propose a novel device in which the DW is driven by the field-like torque with the out-of-plane spin polarization. We first produce the Neel and Bloch wall by changing the sample width. The current is then applied to drive the DW motion. Although the results show that both types of DWs can be moved, they are easily distorted since its profile is supported by the weak demagnetizing field. The resulting $v_{DW}$ is only 36 m/s. Next, an external field is applied to stabilize the DW during dynamics, which gives rise to a much faster $v_{DW}$ of 126 m/s. The external field is then replaced by the DMI to achieve a field-free system and the maximum $v_{DW}$ obtained using our parameters is of 506 m/s. Therefore, in our proposed device, current driven efficient DW motion can be achieved in any types of DWs under zero field. This work provides new degrees of freedom in achieving fast DW motion and could stimulate further works in the mechanism study and device applications.

**References**

**Corresponding Author:** †zhuzhf@shanghaitech.edu.cn

**Acknowledgements**

This work was supported by Shanghai Sailing Program (Grant No. 20YF1430400) and National Science Foundation of China (Grants No. 62074099).

**Figures**

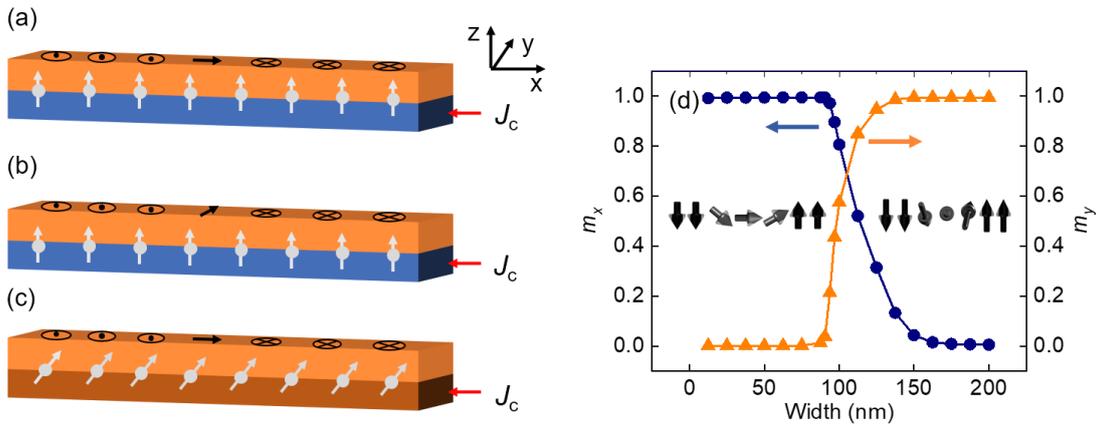

**Figure 1**. Illustrations of the current driven DW motion in systems with perpendicular magnetization. The current induced spin accumulation is marked at the interface. Neel wall (a) and Bloch wall (b) with $\sigma_z$, where the bottom layer can be materials with reduced symmetry. (c) Neel wall with $\sigma_y$, where the bottom layer can be heavy metals. (d) The magnetization at the DW center as a function of sample width. Inset illustrates the profile of Neel wall (left) and Bloch wall (right).

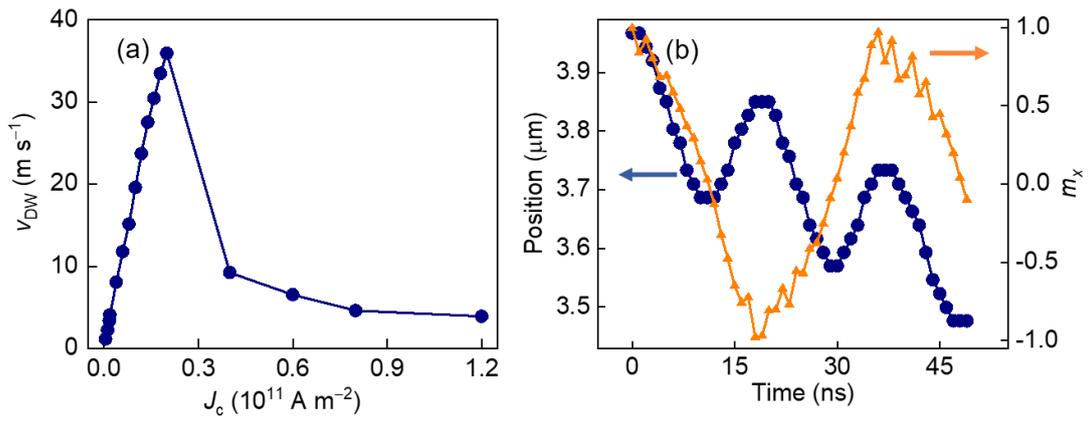

**Figure 2**. (a) Domain wall velocity as a function of current density in the Neel wall. Walker breakdown appears at $J_c = 0.2\times10^{11}$ A/m$^2$. (b) The DW position (left $y$ axis) and $m_x$ (right $y$ axis) as a function of time at $J_c = 0.6\times10^{11}$ A/m$^2$.

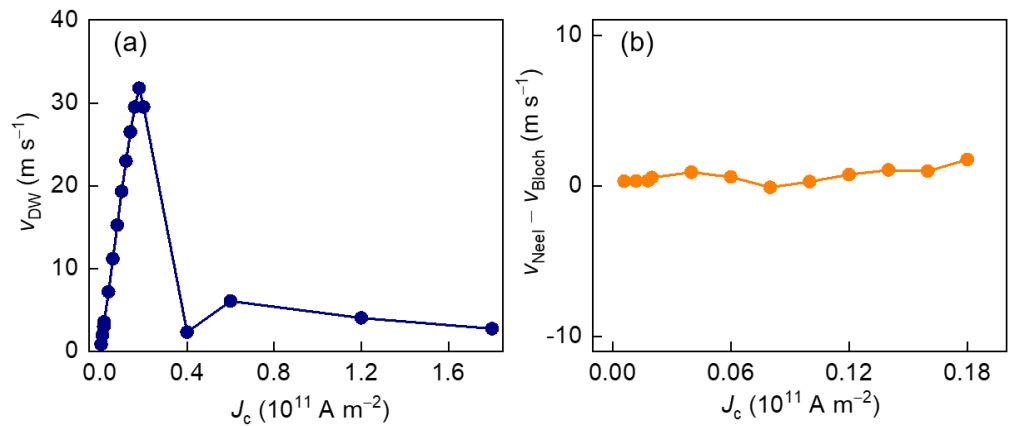

**Figure 3**. (a) Domain wall velocity as a function of current density in the Bloch wall. (b) The difference in the domain wall velocity between Neel wall and Bloch wall in the linear region, i.e., $J_c < 0.18\times10^{11}$ A/m$^2$.

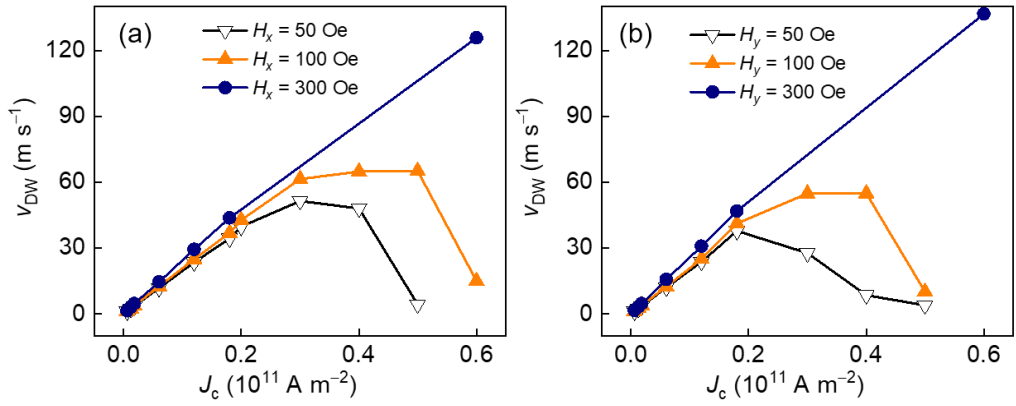

**Figure 4**. The domain wall velocity as a function of current density under different external field in the (a) Neel wall and (b) Bloch wall. $H$ = 50 Oe, 100 Oe and 300 Oe are represented by empty down-pointing triangle, filled up-pointing triangle, and filled circle, respectively.

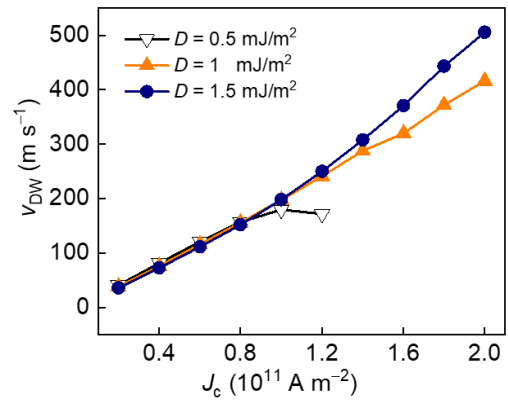

**Figure 5**. Domain wall velocity as a function of current density for samples with different DMI strengths. $D$ = 0.5 mJ/m$^2$, 1.5 mJ/m$^2$ and 2.0 mJ/m$^2$ are represented by empty down-pointing triangle, filled up-pointing triangle and filled circle, respectively. For $D$ = 0.5 mJ/m$^2$, the Walker breakdown appears at $J_c$ = 1 × 10$^{11}$. No Walker breakdown is observed for the other two cases.

Table 1
Summary of domain wall velocity

| Conditions | $v_{DW}$ (m/s) |
|---|---|
| $H_x$ = 0, $D$ =0, Neel wall | 36 |
| $H_x$ = 300 Oe, $D$ =0, Neel wall | 126 |
| $H_x$ = 0, $D$ = 1.5 mJ/m$^2$, Neel wall | 506 |